\documentclass[showpacs,reprint,amsmath,amssymb,aps,prl]{revtex4-1}

\usepackage{graphicx}
\usepackage{dcolumn}
\usepackage{bm}
\usepackage{subfigure}
\usepackage{natbib}
\newcommand{\I}{\mathrm{i}}
\newcommand{\E}{\mathrm{e}}
\newcommand{\D}{\mathrm{d}}

\begin{document}

\preprint{Ptychography}

\title{Time-domain ptychography}

\author{Dirk Spangenberg}
\author{Pieter Neethling}
\author{Erich Rohwer}
\affiliation{Laser Research Institute, Stellenbosch University, Private Bag X1, 7602 Matieland, South Africa
}

\author{Michael H. Br\"ugmann}
\author{Thomas Feurer}
\affiliation{Institute of Applied Physics, University of Bern, Sidlerstrasse 5, 3012 Bern, Switzerland}

\date{\today}

\begin{abstract}
Through dedicated measurements in the optical regime we demonstrate that ptychography can be applied to reconstruct complex-valued object functions that vary with time from a sequence of spectral measurements. A probe pulse of approximately 1~ps duration, time delayed in increments of 0.25~ps is shown to recover dynamics on a ten times faster time scale with an experimental limit of approximately 5~fs. 
\end{abstract}

\maketitle



X-ray diffraction imaging is a promising concept for realizing lens-less imaging of aperiodic objects with atomic-scale resolution. The key challenge is to reconstruct the phase of the diffracted wave and several solutions to this so-called phase problem have been demonstrated \cite{GerchbergSaxton1972, Fienup1982, Miao_et_al1999, Nugent_et_al2003, Eisebitt_et_al2004, Chapman_et_al2006, MaidenRodenburg2009, Greenbaum_et_al2012, Marrison_et_al2013, Fienup2013}. One of the most robust techniques is ptychography. Its concept is related to the solution of the phase problem in crystallography as proposed by Hoppe \cite{Hoppe1969}. It was first demonstrated at optical wavelengths \cite{Rodenburg_et_al2007}, but its scientific impact is expected to be largest at x-ray wavelengths, especially after the commissioning of high brightness coherent x-ray sources. In ptychography the real space image of an object is reconstructed iteratively from a series of far-field diffraction patterns. Each pattern is recorded after either moving the object or the coherent illumination beam in a plane perpendicular to the propagation direction of the illumination beam. The transverse shift of the illumination beam is smaller than its spatial support, so that subsequent patterns result from different, but overlapping regions of the real space object. This ensures that the phase can be extracted. Moreover, the redundant information from overlapping regions helps to improve the convergence of the iterative algorithms. The spatial resolution is limited by the positioning accuracy, the stability of the entire setup, and by the angular range of scattered wavevectors that can be recorded with a sufficiently high signal-to-noise ratio.

So far, x-ray ptychography has been successfully applied in reconstructing real-space objects of up to three spatial dimensions and a spatial resolution down to 16~nm has been demonstrated \cite{Holler_et_al2014}. Here, we propose to extend ptychography by including the temporal dimension and thus to facilitate the reconstruction of spatiotemporal objects, i.e., one-, two-, or three-dimensional objects whose shape or structure varies with time. Far-field diffraction measurements combined with spectral measurements should allow for reconstructing real-time movies of aperiodic atomic-scale objects. The accessible time scales are determined mostly by the range of the spectral measurements, and with present-day high brightness coherent XUV and x-ray sources the time resolution could potentially reach the attosecond regime.

\begin{figure}[htb]
\centering \includegraphics[width=\columnwidth]{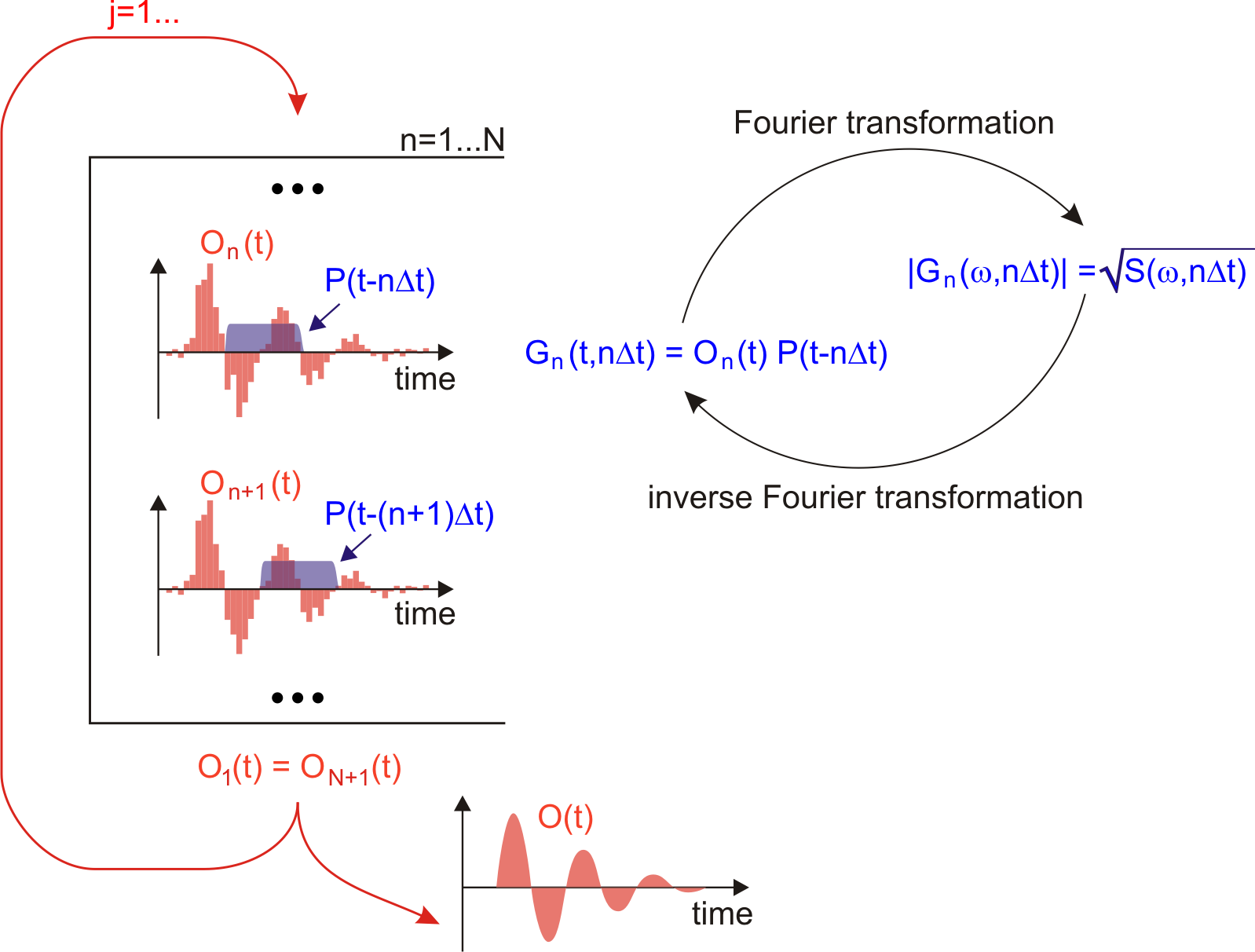}
\caption{Ptychographic iterative scheme for the reconstruction of a complex-valued, time-varying object function $O(t)$ from a sequence of $N$ spectra $S(\omega,n \Delta t)$ recorded at probe pulse time delays $n \Delta t$ with $n \in \{1, \ldots, N\}$. After all $N$ spectra have been processed, the first iteration ($j=1$) is complete and the next iteration ($j=2$) will start until convergence is reached. For all $n$ the top arrow implies that the product of the estimated object function [$O_n(t)$ in red] with the time delayed probe pulse [$P(t-n \Delta t)$ in blue] is Fourier transformed to the spectral domain. Then, the modulus of the Fourier transform is replaced by the square root of the measured spectral intensity. The result is Fourier transformed back to the time domain and is used to update the estimated object function.}
\label{fig_scheme}
\end{figure}

Here, our main goal is to demonstrate that the ptychographic scheme is able to reconstruct an ``object" uniform in space but varying with time. The ``object'' is illuminated with a sequence of partially overlapping, time-delayed coherent probe pulses, and for each time delay a far-field diffraction pattern, i.e., wavelength-resolved spectrum, is recorded. As in the spatial analog, the exit field is formally described by a product of the object and the probe pulse. Time-domain ptychography then yields information on the ``object'' on time scales much shorter than the duration of the probe pulse.

In general, the one-dimensional phase retrieval problem is ambiguous and different solutions may result in the same far-field intensity measurement. In the framework of ptychography \cite{McCallumRodenburg1992}, however, the uniqueness of the solution in one and two dimensions is warranted as long as the illumination pattern is known. The problem of uniqueness, specifically in the one-dimensional case, is also discussed in Ref.~\cite{GuizarSicairos_et_al2010}.  

The basic algorithm, known as the ptychographic iterative engine (PIE) \cite{Faulkner2005}, schematically depicted in Fig.~\ref{fig_scheme}, starts with a random, complex-valued object function. In every iteration $j$ all measured spectra ($n=1 \ldots N$) are processed. The algorithm calculates the exit field $G_n(t,n \Delta t)$ for a particular time delay $n \Delta t$ of the probe pulse and the current estimate of the object function $O_n(t)$

\begin{equation}
G_n(t, n \Delta t) = O_n(t) \; P(t-n \Delta t),
\end{equation}

where $n \in \{1, \ldots, N\}$ indicates one of the measured intensity spectra. From $G_n(t, n \Delta t)$ we calculate the Fourier transform $G_n(\omega,n \Delta t)$ and replace its modulus by the square root of the measured spectrum $S(\omega,n \Delta t)$ while preserving its phase. After an inverse Fourier transformation, the new function $G'_n(t, n \Delta t)$ differs from the initial estimate and the difference is used to update the current estimate of the object function,

\begin{eqnarray}
\label{eq_pty_iter}
\nonumber
O_{n+1}(t) & = & O_n(t) + \beta \; U(t-n \Delta t) \\
&& \times [G'_n(t,n \Delta t) - G_n(t,n \Delta t)],
\end{eqnarray}

with the weight or window function,

\begin{equation}
\label{eq_pty_wt}
U(t) = \frac{|P(t)|}{\mathrm{max}(|P(t)|)} \; \frac{P^*(t)}{|P(t)|^2+\alpha},
\end{equation}

and the two constants $\alpha < 1$ and $\beta \in \; ]0 \ldots 1]$. The choice of $\alpha$ is determined mostly by the noise level and $\beta$ by the probe pulse duration and the time delay $\Delta t$. The best approximation to the actual object function appears typically after only a few iterations $j$ under ideal conditions.

In order to prove the concept and to explore ultrafast time scales we refer to automated pulse shaping of femtosecond light pulses \cite{Vaughan2004}. This methodology allows one to modulate the spectrum of a femtosecond light pulse and thus to tailor its temporal intensity. We will employ it to generate different object functions on the one hand and suitably time delayed probe pulses on the other hand. This approach has the advantage that a variety of different object functions can be programed and the reconstructed results can be readily evaluated. Note that here the object function as well as the probe pulse are conveniently described by slowly varying field envelopes $O(t)$ and $P(t)$ modulating the baseband frequency $\omega_\mathrm{p}$. As a result, $G_n(t, n \Delta t)$ is a slowly varying envelope with a baseband frequency of $\omega_\mathrm{g} = 2 \omega_\mathrm{p}$, around which the measured spectra will be centered.


\begin{figure}[htb]
\begin{center}
\includegraphics[width=0.9\columnwidth]{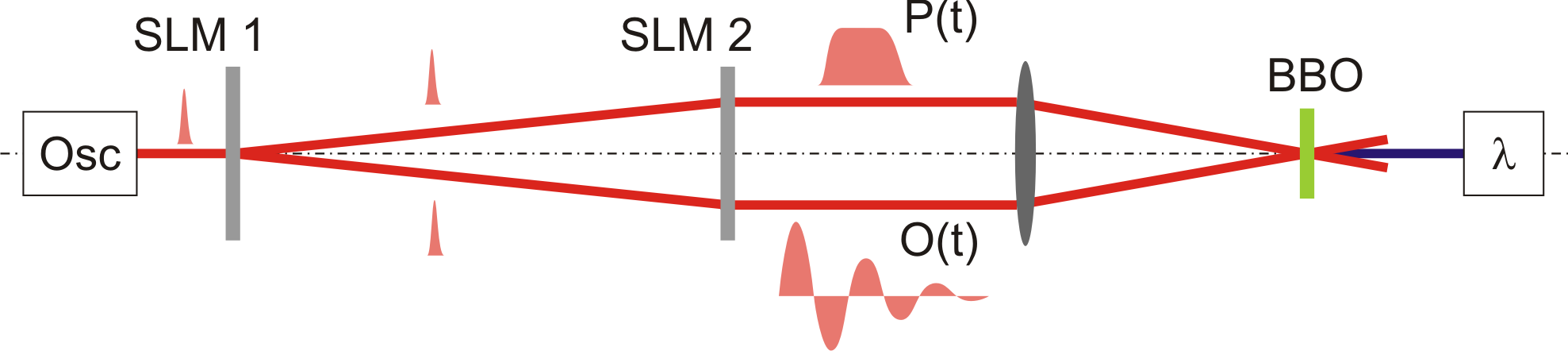}
\caption{Schematic illustrating the concept of the experimental setup.}
\label{fig_setup}
\end{center}
\end{figure}

A schematic of the experimental setup is shown in Fig.~\ref{fig_setup}. The pulse source is an 80~MHz Ti:sapphire oscillator which delivers 80~fs pulses centered at 800~nm. The first two-dimensional spatial light modulator (SLM1) is loaded with a binary hologram to diffract the incoming beam into the plus first and the minus first diffraction orders. Both beams are independently shaped in time by a pulse-shaping apparatus, which includes a second two-dimensional spatial light modulator (SLM2) \cite{Vaughan2004}, and are subsequently focused to a 100-$\mu$m-thick beta barium borate crystal where the exit field, i.e., the product field, is produced through sum-frequency generation. The resulting spectra centered at 400~nm are analyzed by a spectrometer covering the range of 300-545~nm with a resolution of 0.18~nm.

We programed a ``long'' probe pulse by selecting a spectral slice of width $\Delta\Omega$ centered at 800~nm from the source pulse spectrum $E_0(\Omega)$, i.e., $P(\Omega) = M_\mathrm{p}(\Omega) \; E_0(\Omega)$, with

\begin{equation}
\label{eq_probe}
M_\mathrm{p}(\Omega) = \left\{
\begin{array}{rcl}
\E^{-\I \Omega n \Delta t} & \mathrm{for} \;\; | \Omega | \leq \frac{\Delta\Omega}{2}, \\
0 & \mathrm{otherwise},
\end{array}
\right.
\end{equation} 

and the relative frequency $\Omega=\omega-\omega_\mathrm{p}$. The time delay is realized by a linear spectral phase, i.e., $\E^{-\I \Omega n \Delta t}$. The pulse-shaping apparatus (SLM2) allows for time delays within a $\pm 5$~ps time window with a measured resolution of 0.5~as. For sufficiently thin slices, i.e., for $\Delta\Omega \ll \Delta\omega$, where $\Delta\omega$ is the spectral width of the source pulse, the probe pulse is well approximated by a sinc-shaped slowly varying envelope.

The object function $O(t)$ is generated by applying an appropriately chosen transfer function $M_\mathrm{o}(\Omega)$ to the SLM2, i.e., $O(\Omega) = M_\mathrm{o}(\Omega) \; E_0(\Omega)$. We have applied a number of different transfer functions, such as polynomial phases, sinusoidal phases, cosinusoidal amplitudes, and more complex transfer functions. The transfer functions discussed here are given by

\begin{eqnarray}
\label{eq_pulses1}
M_\mathrm{o1}(\Omega) & = & \frac{1-\E^{-\gamma}}{1-\E^{-\gamma (M+1)}} \sum\limits_{m=0}^M (\pm 1)^m \E^{-\I m \Omega \tau - \gamma m}, \\
\label{eq_pulses2}
M_\mathrm{o2}(\Omega) & = & \E^{\I A \sin(\Omega \tau)},
\end{eqnarray}

where the first applies a combined amplitude and phase and the second a pure phase modulation. The first transfer function produces a train of exponentially decaying pulses with $M$ determining the number of pulses, $\gamma$ their exponential decay, and $\tau$ their temporal separation. The plus (minus) sign produces a unipolar (alternating) train of pulses. The sinusoidal phase modulation is determined by the amplitude $A$ and the time delay $\tau$.


\begin{figure}[htb]
\begin{center}
\includegraphics[width=0.9\columnwidth]{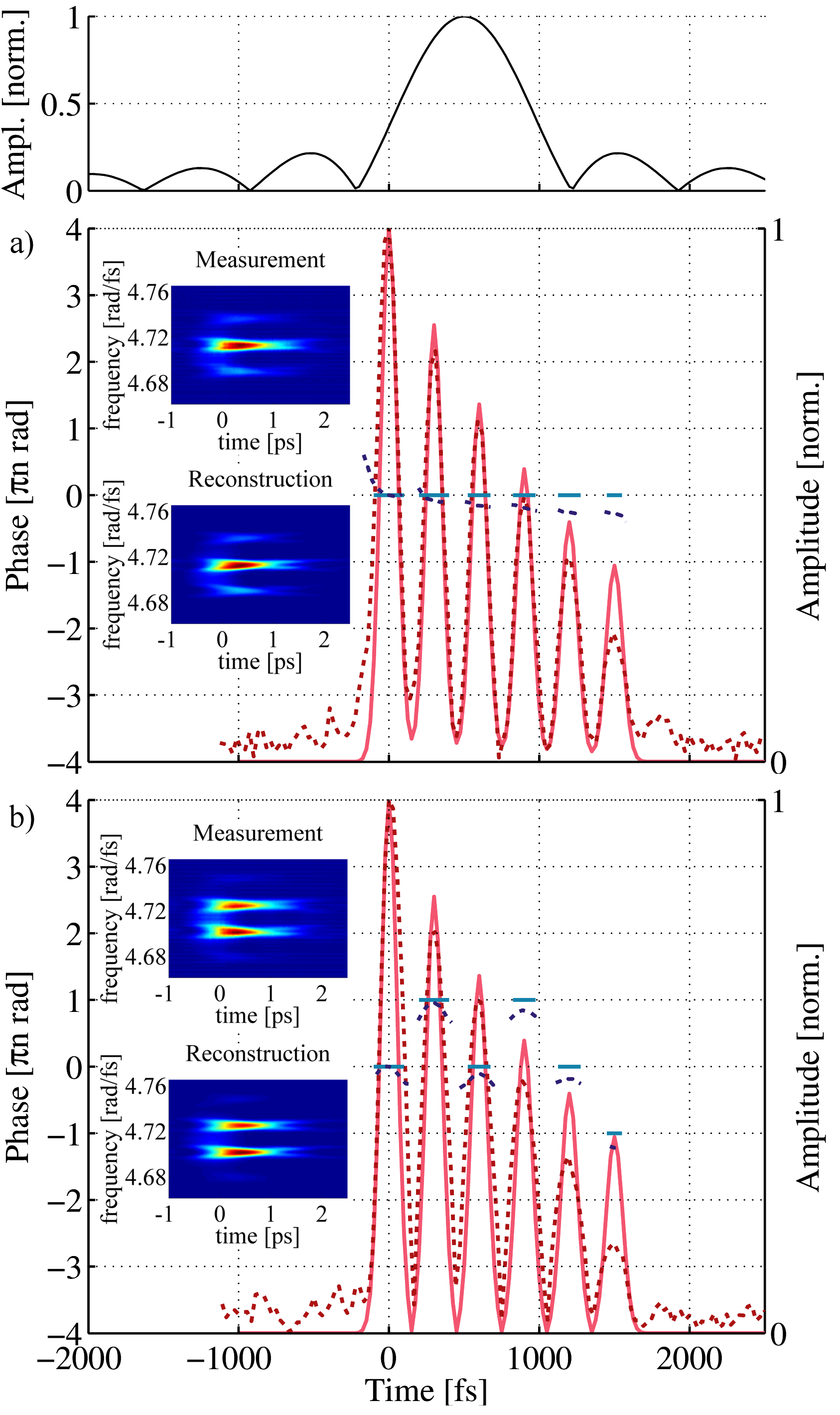}
\caption{Theoretical (solid curves) and reconstructed (dashed curves) amplitude and phase of the object function defined in Eq.~\eqref{eq_pulses1} with (a)the plus sign and (b) the minus sign. For convenience, the phase is only shown in regions where the amplitude is larger than 0.1. The top panel shows the probe pulse $P(t)$ with a duration of approximately 900~fs. The corresponding parameters are $\tau=300$~fs, $M=5$, and $\gamma=1/5$.}
\label{fig_ampl}
\end{center}
\end{figure}

\begin{figure}[htb]
\begin{center}
\includegraphics[width=0.9\columnwidth]{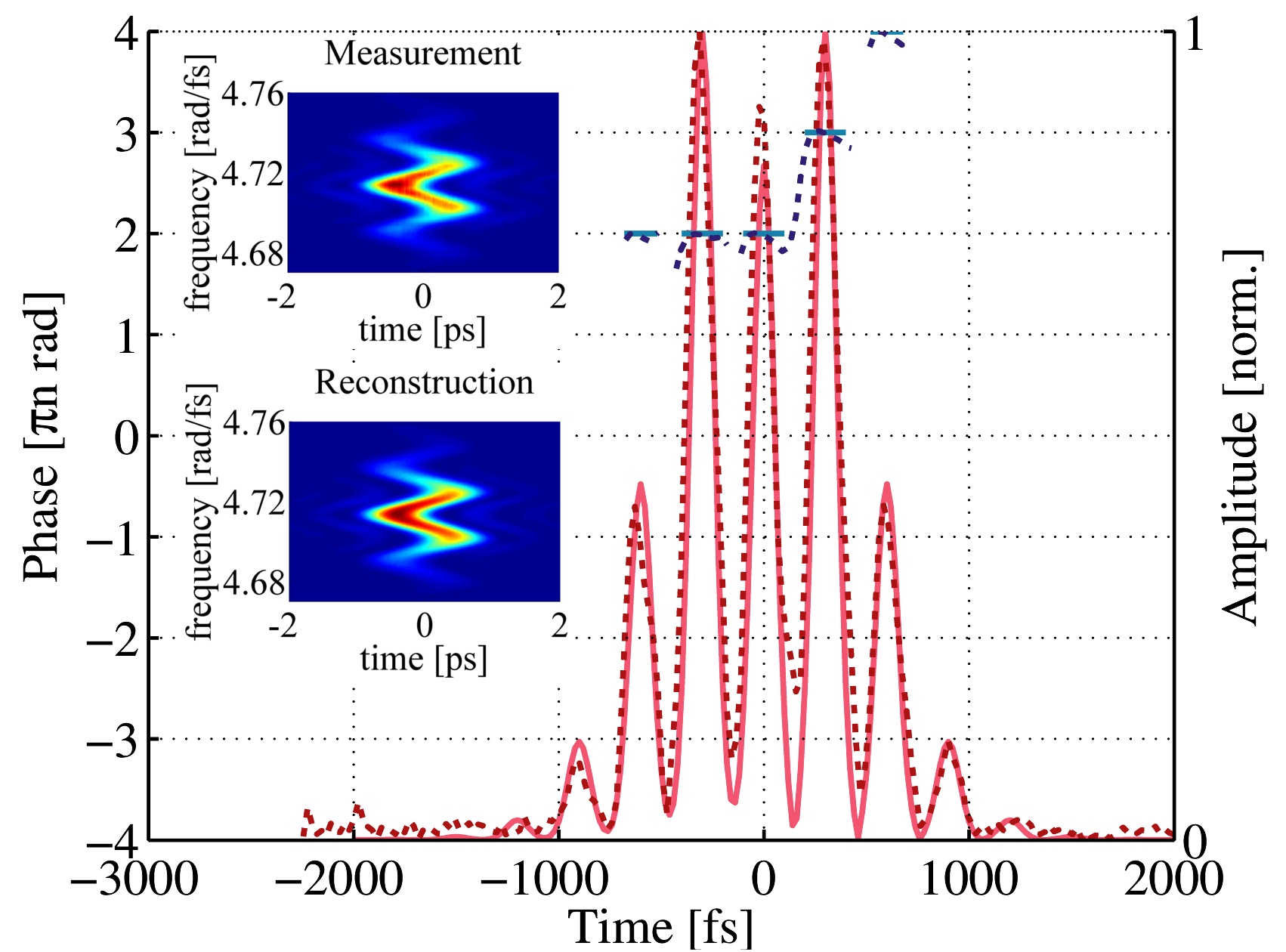}
\caption{Theoretical (solid curves) and reconstructed (dashed curves) amplitude and phase of the object function defined in Eq.~\eqref{eq_pulses2}. The corresponding parameters are $A=1.57$~rad and $\tau=300$~fs.}
\label{fig_sin}
\end{center}
\end{figure}

The experimental results as well as the reconstructions are presented in Figs.~\ref{fig_ampl} and~\ref{fig_sin}. For the probe pulse a spectral window of 3~nm was used, resulting in a sinclike pulse $P(t)$ with a duration of approximately 900~fs. The parameters of the first object function in Eq.~\eqref{eq_pulses1} were $\tau=300$~fs, $M=5$, and $\gamma=1/5$ and for Fig.~\ref{fig_ampl} a) the ``plus sign'' and for Fig.~\ref{fig_ampl} b) the ``minus sign'' in Eq.~\eqref{eq_pulses1} was selected. The time-frequency distributions show the measured (top) and the simulated (bottom) spectra as a function of time delay with an increment of 25~fs. Only a subset of all measured spectra is used for the reconstruction, e.g. every tenth spectrum corresponding to a $\Delta t$ of 250~fs. The theoretical object functions, as defined in Eq.~\eqref{eq_pulses1}, are indicated by the solid curves versus time, i.e., the amplitude (solid red curve) and the phase (solid blue curve). For convenience, the phase is only shown in regions where the amplitude is larger than 0.1. The parameters of the update function were $\alpha=0.3$, $\beta=0.5$, and the sampling rate was exact Nyquist sampling. A suitable value for $\alpha$ was determined by analyzing the rms error (rms refers to the root-mean-square difference between theoretical and reconstructed spectrogram) for a set of simulated noisy spectra as function of $\alpha \in[0,1]$ and the signal-to-noise ratio (SNR). For a measured SNR of $>500$, $\alpha \approx 0.3$ leads to the smallest rms. Similarly, a suitable value for $\beta$ was determined by analyzing the rms error as a function of $\beta \in [0,1]$ and the ratio of $\Delta t$ and the probe pulse duration. We would like to emphasize thatalthough $\alpha$ as well as $\beta$ can be varied within a wide range of values, the final result varies very little, however, the rate of convergence may slow down by a factor of up to 10.

The parameters for the object function in Eq.~\eqref{eq_pulses2} were $A=1.57$~rad and $\tau=300$~fs. Figure~\ref{fig_sin} shows that the sinusoidal spectral phase modulation produces a sequence of temporal diffraction orders whose amplitude and phase are well reproduced by the ptychographic reconstruction algorithm.

Convergence in the reconstruction is typically observed after several tens of iterations, and the results shown in Figs.~\ref{fig_ampl} and~\ref{fig_sin} are the asymptotic object functions after 900 iterations. In all three cases the reconstructed amplitudes as well as the phases agree very well with the target object functions. Moreover, comparing the reconstructed phases in Figs.~\ref{fig_ampl} a), \ref{fig_ampl} b) and~\ref{fig_sin} confirms that the temporal phase, which is a result of the iterative algorithm, is recovered with a good degree of accuracy. We do, however, find small deviations between the reconstructed and the theoretical object functions, e.g., in Fig.~\ref{fig_ampl}, which are due to a number of experimental uncertainties. For example, the probe pulse used in the reconstruction process differs somewhat from the true probe pulse due to a number of reasons which are related to the pulse-shaping apparatus itself (details can be found in Ref.~\cite{Vaughan2004}). Briefly, the pulse shaper approximates the spectral transfer function via the discrete pixels of the spatial light modulator, that is, the spectral transfer function applied consists of discrete steps which are separated by small inactive gap regions. Additionally, the spectral transfer function has to be convoluted with the spectral resolution of the grating lens combination. For all reconstructions in this contribution we have assumed perfect spectral resolution, continuous transfer functions, and we have neglected the gaps, which explains at least part of the deviations observed. However, we believe that the quality of the reconstruction is sufficient to prove that ptychography can be applied to reconstruct ultrafast time-dependent object functions.


The results demonstrate that the ptychographic iterative scheme can extract a time-varying, complex-valued object function from a sequence of spectra each recorded for a different time-delayed coherent probe pulse, and we believe that we have shown sufficient experimental evidence to prove that the concept of ptychography is applicable to ultrafast lensless time-domain imaging. With an approximately 1-ps-long probe pulse, time-delayed in increments of 250~fs, it was possible to reconstruct temporal features with a time constant of about 100~fs, i.e., temporal variations ten times shorter than the duration of the probe pulse. Given the spectral range of the spectrometer (300-545~nm), the fastest detectable temporal feature is approximately 5~fs. This is comparable to ptychography experiments in the spatial domain, e.g., in Ref.~\cite{Holler_et_al2014}, a 700-nm-wide beam was used to reconstruct objects with a 11~nm resolution. The accuracy with which the time delay can be adjusted through pulse shaping is exceptionally high, here 0.5~as, and thus much better than the accuracy of most spatial analogs. There are several applications of this modality, some of which which we will pursue.

With the experimental parameters chosen appropriately, the scheme may also prove to be useful for ultrashort pulse characterization purposes. The object function is identified with the electric field of the unknown pulse and the probe pulse with a spectrally filtered copy of it or another suitably prepared probe pulse. A related and widely used pulse characterization scheme is frequency-resolved optical gating (FROG), in which a spectrogram is measured and an inversion algorithm is used to retrieve the object function, i.e., the unknown pulse \cite{Kane2008}. The ptychographic scheme, however, differs from FROG in several ways. For example, the time step is not linked directly to sampling and the temporal resolution is dominated by the largest frequency shift detectable rather than by the sampling of the spectrogram. 

Also, the methodology is envisaged to be valuable for time-resolved pump-probe spectroscopy schemes. Typically, an ultrashort pump pulse triggers a material response (object function) which is then probed by an ultrashort probe pulse. Irrespective of the specific type of spectroscopy, the probe pulse duration in most schemes is shorter than the fastest dynamical feature to be measured. With the ptychographic probing scheme the requirement of a sufficiently short probe pulse is obviously relaxed. In nonlinear spectroscopy, the signal is either measured directly (homodyne) or through spectral interferometry with a local oscillator (heterodyne). Since a heterodyne measurement yields the amplitude and phase of the nonlinear signal, the ptychographic scheme is useful only in homodyne measurements. Therefore, time-domain ptychography is advantageous if the homodyne signal is of sufficient strength, because less data have to be recorded, or if sufficiently short pulses for standard pump-probe spectroscopy (which includes heterodyne detection) are not readily available, for example, in the UV. Another advantage, albeit more technical in nature, is that beam delivery systems no longer need to be extremely broadband, e.g., dielectric mirrors, wave plates, etc., which are often expensive and difficult to produce. For example, in homodyne transient grating spectroscopy the generated third order polarization can be expressed as

\begin{equation}
\label{eq_TG}
\mathcal{P}^{(3)}(\vec{k},t) = -P(t) \; \int\limits_0^\infty \D t_1 \; \chi^{(3)}(t_1) \; \left| E_\mathrm{p}(t+\tau-t_1) \right|^2, 
\end{equation}

with the time delay $\tau$ between the pump pulse $E_\mathrm{p}(t)$ forming the transient grating and the probe pulse $P(t)$. Ptychography would allow one to reconstruct the convolution of the pump pulse and the third-order material response without the need of a short probe pulse. If the pump pulse is known from a separate measurement or assumed to be impulsive, the material response can be readily extracted. With the emergence of spatially and temporally coherent pulsed x-ray sources the scheme proposed here may also be applicable to optical-pump-x-ray probe and x-ray pump-x-ray probe schemes.

\begin{acknowledgments}
We gratefully acknowledge fruitful discussions with A.~Cannizzo, B.~Patterson, and M.~Guizar-Sicairos and financial support from the CSIR NLC and the NCCR MUST research instrument of the Swiss National Science Foundation.
\end{acknowledgments}


\end{document}